\documentclass[aps,prb,twocolumn,floatfix,superscriptaddress]{revtex4-2}
\usepackage{graphicx,amsfonts,bm,color}
\usepackage[normalem]{ulem}

\begin{document}

\title{Microscopic origins and stability of the ferromagnetism in Co$_3$Sn$_2$S$_2$}

\author{I. V. Solovyev}
\email{SOLOVYEV.Igor@nims.go.jp}
\affiliation{National Institute for Materials Science, MANA, 1-1 Namiki, Tsukuba,
Ibaraki 305-0044, Japan}
\affiliation{Department of Theoretical Physics and Applied Mathematics, Ural Federal University,
Mira str. 19, 620002 Ekaterinburg, Russia}
\affiliation{Institute of Metal Physics, S. Kovalevskaya str. 18, 620108 Ekaterinburg, Russia}
\author{S. A. Nikolaev}
\affiliation{National Institute for Materials Science, MANA, 1-1 Namiki, Tsukuba,
Ibaraki 305-0044, Japan}
\affiliation{Tokyo Tech World Research Hub Initiative (WRHI), Institute of Innovative Research, Tokyo Institute of Technology, 4259 Nagatsuta, Midori-Ku, Yokohama, Kanagawa, 226-8503, Japan}
\affiliation{Laboratory for Materials and Structures, Tokyo Institute of Technology, 4259 Nagatsuta, Midori-Ku, Yokohama, Kanagawa, 226-8503, Japan}
\author{A. V. Ushakov}
\affiliation{Institute of Metal Physics, S. Kovalevskaya str. 18, 620108 Ekaterinburg, Russia}
\author{V. Yu. Irkhin}
\affiliation{Institute of Metal Physics, S. Kovalevskaya str. 18, 620108 Ekaterinburg, Russia}
\author{A. Tanaka}
\affiliation{National Institute for Materials Science, MANA, 1-1 Namiki, Tsukuba,
Ibaraki 305-0044, Japan}
\author{S. V. Streltsov}
\affiliation{Department of Theoretical Physics and Applied Mathematics, Ural Federal University,
Mira str. 19, 620002 Ekaterinburg, Russia}
\affiliation{Institute of Metal Physics, S. Kovalevskaya str. 18, 620108 Ekaterinburg, Russia}

\date{\today}

\date{\today}
\begin{abstract}
Based on the density functional theory, we examine the origin of ferromagnetism in the Weyl semimetal Co$_3$Sn$_2$S$_2$ using different types of response theories. We argue that the magnetism of Co$_3$Sn$_2$S$_2$ has a dual nature and bears certain aspects of both itineracy and localization. On the one hand, the magnetism is soft, where the local magnetic moments strongly depend on temperature and the angles formed by these moments at different Co sites of the kagome lattice, as expected for itinerant magnets. On the other hand, the picture of localized spins still remains adequate for the description of the local stability of the ferromagnetic (FM) order with respect to the transversal spin fluctuations. For the latter purposes, we employ two approaches, which provide quite different pictures for interatomic exchange interactions: the regular magnetic force theorem and a formally exact theory based on the calculation of the inverse response function. The exact theory predicts Co$_3$Sn$_2$S$_2$ to be a three-dimensional ferromagnet with the strongest interaction operating between next-nearest neighbors in the adjacent kagome planes. The ligand states are found to play a very important role by additionally stabilizing the FM order. When the local moments decrease, the interplane interactions sharply decrease, first causing the FM order to become quasi-two-dimensional, and then making it unstable with respect to the spin-spiral order propagating perpendicular to the kagome plane. The latter instability is partly contributed by the states at the Fermi surface and may be relevant to the magnetic behavior of Co$_3$Sn$_2$S$_2$ near the Curie temperature. Peculiarities of the half-metallic ferromagnetism in Co$_3$Sn$_2$S$_2$ are also discussed.

\end{abstract}

\maketitle

\section{\label{sec:Intro} Introduction}
\par The shandite Co$_3$Sn$_2$S$_2$, hosting the kagome lattice of Co ions, is a fascinating magnetic material, in several respects. It is a ferromagnet with small spontaneous magnetization (about $0.3$ $\mu_{\rm B}$ per Co site), but a relatively high Curie temperature $T_{\rm C} \sim 177$ K. It has attracted a lot of attention as a magnetic Weyl semimetal those non-trivial topology of the electronic states gives rise to a large anomalous Hall effect (AHE)~\cite{Liu,Xu,Minami,Yanagi}. The quantum AHE was also realized by fabricating the two-dimensional devices on the basis of Co$_3$Sn$_2$S$_2$~\cite{Tanaka,Muechler}. These intriguing magnetic properties are further amplified by the fact that Co$_3$Sn$_2$S$_2$ is half-metalic~\cite{Jiao}, where the conduction takes place only in one spin channel while another spin channel is gapped. This half-metallicity remains robust upon doping in Co$_3$In$_x$Sn$_{2-x}$S$_2$, where the ground-state magnetization depends linearly on $x$ and persists up to $x \simeq 0.85$, as was demonstrated in theoretical~\cite{Yanagi} and experimental~\cite{Zhou,Kassem} studies. The coexistence of the easy-axis ferromagnetic (FM) and planar $120^{\circ}$ antiferromagnetic (AFM) orders was proposed from the $\mu$SR measurements in the temperature range $T_{\rm A} < T < T_{\rm C}$ (where $T_{\rm A} \sim 90$ K), and the intensity of the AHE was proportional to the fraction of the FM phase~\cite{Guguchia.NatComm}. The anomaly of the magnetic susceptibility, which may be relevant to the two-phase state, was also observed but at somewhat higher $T_{\rm A} \sim 130$ K~\cite{ZhangPRL}. On the other hand, no evidence of the AFM component up to $T_{\rm C} \sim 177$ K was observed in recent unpolarized neutron diffraction and spherical neutron polarimetry measurements~\cite{SohArxiv}.

\par Very generally, the Weyl semimetals require either spatial inversion or time reversal symmetry to be broken. While the early realizations were initially all from the former category, the latter direction attracts more and more attention recently. For instance, the intrinsic AHE in Co$_3$Sn$_2$S$_2$ is associated with the spontaneous time-reversal symmetry breaking caused by the FM order. In this sense, the origin of this FM order is one of the key question in the physics of Co$_3$Sn$_2$S$_2$. Nevertheless, it is not fully understood and remains largely controversial~\cite{SavrasovPRB2021}.

\par The small magnetization is believed to be related to the cluster effects, which also reduce the effective Coulomb interactions~\cite{SavrasovPRB2021}, as expected for molecular type compounds~\cite{lacunar}.

\par The half-metallic state implies the absence of Stoner excitations, so that the important role of spin fluctuations is generally expected in Co$_3 $Sn$_2 $S$ _2 $  (e.g., in the temperature dependence of magnetic moment). Indeed, experimental magnetization curves~\cite{Kassem} for Co$_3$In$_x$Sn$_{2-x}$S$_2$ demonstrate strong fluctuations and are reminiscent of those for weak itinerant ferromagnets, especially at large $x$, where the ground state moment is strongly reduced. The electronic structure is also expected to be unusual and featured by the appearance of non-quasiparticle states in the gap owing to the electron-magnon scattering~\cite{RMP}. On the other hand, with increasing $T$ beyond the spin-wave region, the spin fluctuations inherent to the itinerant magnets should play a role. This is confirmed by the parameters of Takahashi's theory~\cite{Takahashi} obtained from the Arrott plot and fitted to a generalized Rhodes–Wohlfarth plot,  $p_{\rm eff}/p_{\rm s}$ versus $T_{\rm C}/T_{\rm 0}$ ($p_{\rm s}$ being the spontaneous moment, $p_{\rm eff}$ being the effective moment, and $T_{\rm 0}$ being a measure of the spin-fluctuation spectral distribution in the frequency space): $T_{\rm 0} = 1230$ K and $p_{\rm eff}/p_{\rm s}=2.14$ at $x=0$ (and considerably increase with the increase of $x$)~\cite{Kassem}. Such a situation is radically different from half-metallic localized-moment Heusler compounds, where $T_C/T_0 \simeq 1$ and $p_{\rm eff} \simeq p_{\rm s}$~\cite{Sakon}. Besides that, in the quasi-two-dimensional situation, specific fluctuation behavior occurs even in the localized-spin model~\cite{kat07}.

\par The problem of exchange interactions and stability of the FM state was addressed recently on the basis of combined experimental inelastic neutron scattering studies and theoretical calculations in the framework of density functional theory (DFT)~\cite{Liu2021,ZhangPRL}. On the experimental side, it was concluded that the FM order is primarily stabilized by the long-range ``across-hexagon'' interaction in the kagome plane~\cite{ZhangPRL}. One should note, however, that the available experimental spin-wave dispersion data is limited only to the acoustic branch close to the $\Gamma$ point~\cite{Liu2021,ZhangPRL}. Moreover, the experimental picture of interatomic exchange interaction was in sharp contrast with the results of theoretical calculations, predicting the strongest nearest-neighbor interaction to be in the kagome plane~\cite{Liu2021,ZhangPRL}. On the other hand, the theoretical analysis was based on the magnetic force theorem (MFT)~\cite{LKAG1987}, the validity of which is known to be questionable for the itinerant electron systems as it relies on additional approximations. Therefore, more rigorous theoretical methods may be necessary~\cite{BrunoPRL2003,Antropov2006,PRB2021}.

\par In this work, we systematically study the problem of stability of the FM state in Co$_3$Sn$_2$S$_2$ using different kinds of response theories and argue that the magnetism of Co$_3$Sn$_2$S$_2$ has a dual nature. First, we consider the criteria of emergence of the FM state caused by longitudinal fluctuations of the magnetic moments and show that the behavior of Co$_3$Sn$_2$S$_2$ bears certain similarities to the Stoner picture of itinerant magnetism in the sense that the local moments are pretty soft and can easily evolve with temperature (on a reasonable temperature scale) and depending on the angle between them. Nevertheless, the transversal spin fluctuations, relevant to the rotational spin degrees of freedom, are also important and should be rigorously considered in the analysis of stability of the FM state.

\par The article is organized as follows. In Sec.~\ref{sec:DFT} we briefly discuss the details of DFT calculations and summarize the key results, which are important for understanding the origin of the ferromagnetism in Co$_3$Sn$_2$S$_2$. Then, in Sec.~\ref{sec:model}, we deal with the realistic electronic model extracted from DFT in the basis of Wannier functions and capturing the essential ingredients of the electronic structure of Co$_3$Sn$_2$S$_2$ relevant to the magnetism. Particularly, in Sec.~\ref{sec:existence}, we consider the criteria of emergence of the magnetic state from the nonmagnetic one, which explains the main tendencies of DFT calculations. The analysis is similar to the Stoner theory of magnetism~\cite{Stoner}, but generalized to the case of several different atoms in the primitive cell, including the ligand states. Namely, we explicitly show that the magnetic solution exists up to certain critical angles formed by three Co spins in the kagome lattice and collapses to the nonmagnetic state when the angles exceed these critical values. This is clearly different from the Heisenberg picture of magnetism, which would be expected for the localized spins. Nevertheless, the Heisenberg model can be still introduced locally for the description of local stability of the FM state with respect to the transversal spin fluctuations caused by the infinitesimal rotations of spins~\cite{LKAG1987}. We consider such model in Sec.~\ref{sec:Jij}. For these purposes we employ a formally exact theory of interatomic exchange interactions~\cite{PRB2021} and show how it revises the MFT based results. Particularly, the exact theory predicts Co$_3$Sn$_2$S$_2$ to be the three-dimensional ferromagnet with the strongest interaction $J_{5}$ operating between the kagome planes in the fifth coordination sphere. Moreover, the ligand states play a very important role in strengthening the FM interactions. Then, in Sec.~\ref{sec:Jm}, we investigate the dependence of the exchange interactions on the value of total magnetization $M$ in the FM state. By these means we simulate the temperature effects, which according to the Stoner picture should decrease the  magnetization. We show that the inter-plane interactions drastically decrease with the decrease of $M$, making the FM state unstable with respect to the spin-spiral state propagating perpendicular to the kagome planes, which may be relevant to the AFM phase emerging below $T_{\rm C}$~\cite{Guguchia.NatComm}. Finally, in Sec.~\ref{sec:summary}, we summarize our work.

\section{\label{sec:DFT} GGA calculations}

\subsection{\label{sec:DFT_details} Details}
\par First-principles electronic structure calculations for Co$_{3}$Sn$_{2}$S$_{2}$ were performed in the generalized gradient approximation (GGA)~\cite{gga-pbe} for the experimental crystal structure~\cite{structure} using Vienna ab-initio simulation package (VASP)~\cite{vasp} within the framework of projected augmented waves~\cite{paw}. The rhombohedral Brillouin zone was sampled on a mesh of 10$\times$10$\times$10 Monkhorst-Pack $\boldsymbol{k}$-points~\cite{mpack}. The partial occupancies were determined using the Methfessel-Paxton scheme with the smearing of 0.1 eV~\cite{MethfesselPaxton}. The convergence criteria for the total energy calculations was set to 10$^{-8}$ eV. We have considered two types of magnetic structures: (i) a collinear FM state with a fixed value of the total magnetization defined as a difference between the spin-up and spin-down states, (ii) a non-coplanar umbrella-type spin texture (which can be viewed as a continuous transformation of the FM state to the $120^{\circ}$ spin state in the $xy$ plane), by constraining directions of the magnetic moments at three Co sites (while allowing the size of the moments to relax in the course of self-consistency).

\par The electronic structure of Co$_{3}$Sn$_{2}$S$_{2}$ was interpolated in the basis of Wannier functions constructed for the Co 3$d$, Sn 5$p$, and S 4$p$ orbitals using the maximal localization technique~\cite{wannier90}. The calculated band structures were disentangled in the range from $\sim-8$ eV to $\sim5$ eV with respect to the Fermi level, and the states up to $\sim2$ eV above the Fermi level were kept frozen during the wannierization.
\par Nodal lines and positions of the Weyl points were identified based on the Wannier interpolation by using the WannierTools package~\cite{WannierTools}.

\subsection{\label{sec:DFT_summary} Summary of main results}
\par
\begin{figure}[b]
\begin{center}
\includegraphics[width=0.48\textwidth]{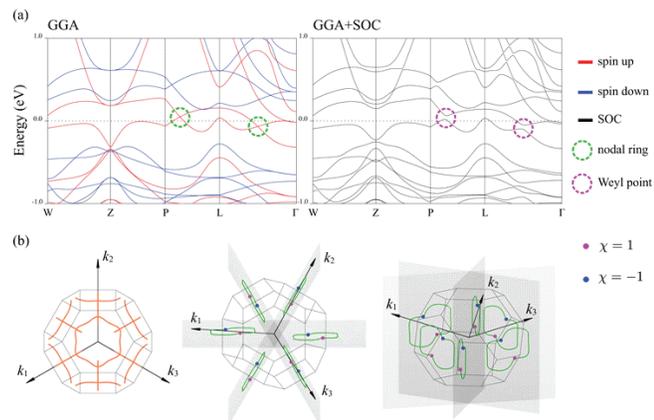}
\end{center}
\caption{(a) Band structure for the collinear ferromagnetic state of Co$_{3}$Sn$_{2}$S$_{2}$ with and without spin-orbit coupling (GGA+SOC and GGA, respectively). (b) Band crossing (left) and nodal lines (middle and right) calculated for the pair of spin-up states at the Fermi level without spin-orbit coupling. Blue and pink points correspond to the Weyl nodes with opposite chiralities, as calculated in the presence of spin-orbit coupling. Grey planes denote the mirror planes in the reciprocal space. Band crossings derived from the gap function are only shown in the first Brillouin zone.}
\label{fig.band_gga}
\end{figure}

 The band structures of Co$_{3}$Sn$_{2}$S$_{2}$ in a collinear FM state calculated without and with spin–orbit coupling are shown in Fig.~\ref{fig.band_gga}.
\noindent
\noindent Co$_{3}$Sn$_{2}$S$_{2}$ is half-metallic where the spin-down channel has a gap of $\sim0.33$~eV at the Fermi level, and the total magnetic moment $M$ is 1 $\mu_{\mathrm{B}}$ per formula unit. Without spin-orbit coupling, the spin-up states in the vicinity of the Fermi level develop linear band crossings along the $\mathrm{P}-\mathrm{L}$ and $\mathrm{L}-\Gamma$ paths due to the band inversion. In fact, the proximity of the spin-up states at the Fermi level and the corresponding gap function $(E_{n+1}-E_{n})^{2}$ has a complicated structure, where the band crossings form closed intersecting lines, as shown in Fig.~\ref{fig.band_gga}(b). Six closed lines lie in the mirror planes of the $D_{3d}$ point symmetry and turn out to be topologically protected in the absence of spin-orbit coupling, forming the nodal lines. In the presence of spin-orbit coupling, the FM state looses its mirror symmetry. This causes the crossings to split and open small gaps with band anti-crossings along the former nodal lines, except for a pair of points for each nodal line where the linear crossing persists. These points known as the Weyl nodes act as a monopole sink and source of the Berry curvature with the opposite topological charges (or chiralities, $\chi=\pm1$).

\par
\begin{figure}[b]
\begin{center}
\includegraphics[width=0.48\textwidth]{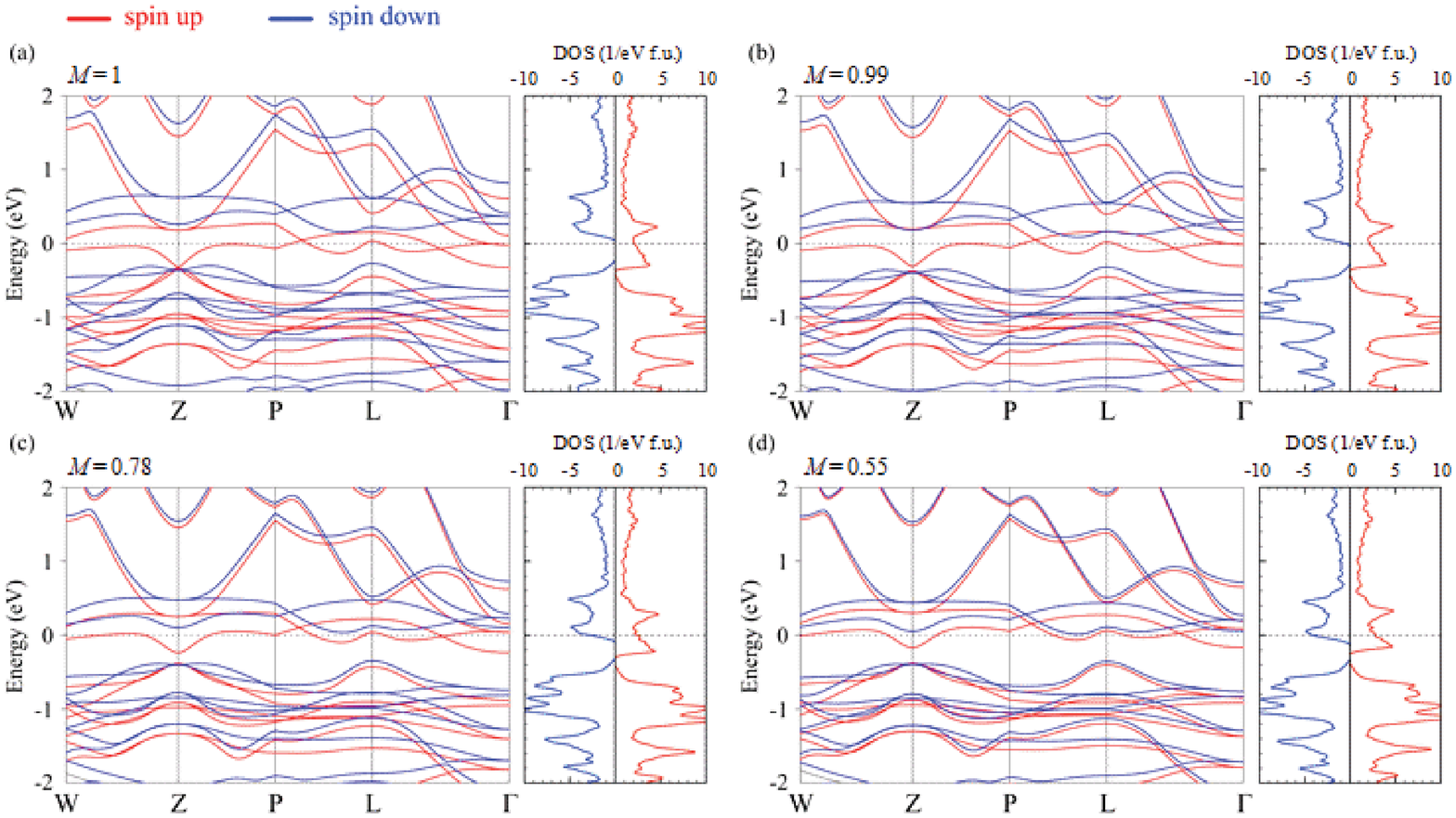}
\end{center}
\caption{Electronic band structures of Co$_{3}$Sn$_{2}$S$_{2}$ in the collinear ferromagnetic state as obtained in constraint calculations with the fixed value of total magnetization: (a) $M=1$ $\mu_{\mathrm{B}}$, (b) $M=0.99$ $\mu_{\mathrm{B}}$, (c) $M=0.78$ $\mu_{\mathrm{B}}$, and (d) $M=0.55$ $\mu_{\mathrm{B}}$. The dashed lines denote the Fermi level.}
\label{fig.band}
\end{figure}

 Deviation of the spin magnetization from the ground state value $M=1$ $\mu_{\mathrm{B}}$ destroys the half-metallic character of the electronic structure, so that the Fermi level crosses the majority spin (spin up) as well as minority spin (spin down) states (Fig.~\ref{fig.band}).
\noindent

\par
\begin{figure}[t]
\begin{center}
\includegraphics[width=0.48\textwidth]{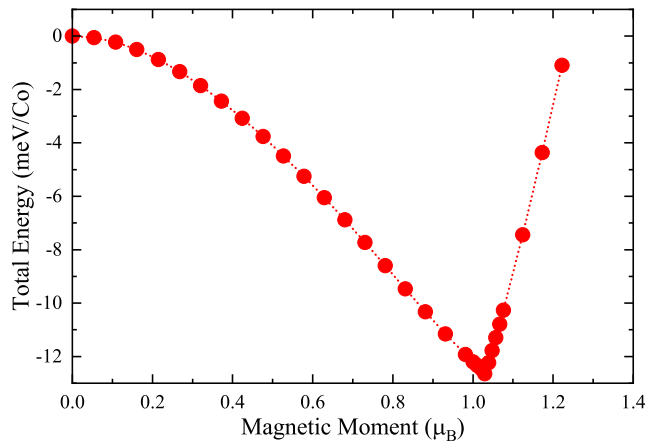}
\end{center}
\caption{
Total energy versus magnetic moment of three Co sites in the unit cell, as obtained in constrained GGA calculations. The magnetic moment was evaluated within atomic spheres of radii $1.3$~\AA. The total energy is calculated relative to the nonmagnetic state.}
\label{fig.EM}
\end{figure}

 Then, let us consider the results of constrained GGA calculations, where we fix the absolute values of magnetic moments at the Co sites in the FM structure.  The dependence of total energy ${\cal E}$ on the magnetic moment is shown in Fig.~\ref{fig.EM}.
\noindent
\noindent The minimum is obtained at $1.035$ $\mu_{B}$ (evaluated with Co atomic spheres of the radii $1.3$~\AA), which corresponds to the total moment $M=1$ $\mu_{B}$ (also including the contributions of the S and Sn sites as well as of the interstitial region). Then, ${\cal E}$ increases with the decrease of $M$. However, the change is relatively small (corresponding to only $143$ K on the temperature scale, which is comparable with the experimental $T_{\rm C}$). This is the first indication of the itinerant character of magnetism in Co$_3$Sn$_2$S$_2$, where the modest elevation of $T$ results in the change of the absolute value of $M$. For instance, a simple thermal averaging with $e^{-{\frac{{\cal E}(M)}{k_{\rm B} T}}}$ will decrease $M$ by about 25\% for temperatures close to $T_{\rm C}$. The derivative discontinuity of the total energy ${\cal E}$ at $M=1$ $\mu_{B}$ is related to the half-metallic character of the electronic structure, where the constraining field, $h= -\frac{\partial {\cal E}}{\partial M}$, required to produce the magnetization in the vicinity of $M=1$ $\mu_{B}$ undergoes a jump of the order of the energy gap in the minority spin channel.

\par This is quite contrary to the expectations based on the band splitting between the majority- and minority-spin states near the Fermi level, being about $0.5$ eV in the ground sate~\cite{SavrasovPRB2021} (see Fig.~\ref{fig.band}). On the temperature scale, this splitting would correspond to $5800$ K and definitely rule out the Stoner picture of magnetism for Co$_3$Sn$_2$S$_2$. However, for the half-metallic state in DFT such splitting is not well defined (the shift of the minority-spin states does not change the energy and magnetization, provided that the Fermi level continues to fall in the gap)~\cite{EschrigPickett}. More generally (and according to the philosophy of DFT), the Kohn-Sham (KS) single particle energies is an auxiliary construction, which does not have a clear physical meaning. Therefore, the thermodynamic properties in DFT should be evaluated using the total energies (instead of the KS single particle ones), which lead to very different temperature scale in the case of Co$_3$Sn$_2$S$_2$.

\par
\begin{figure}[b]
\begin{center}
\includegraphics[width=0.48\textwidth]{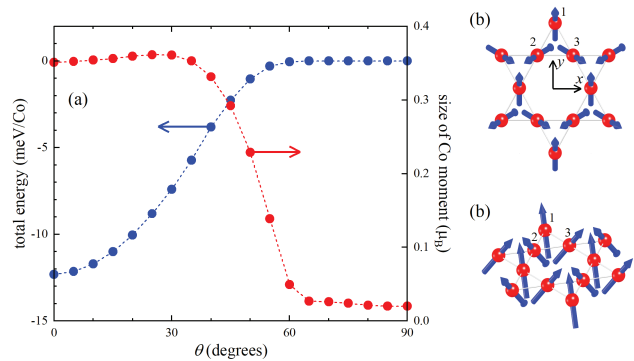}
\end{center}
\caption{
(a) Total energy (left axis $y$) and size of magnetic moment at the Co site (right axis $y$) as obtained in the constrained GGA calculations for the umbrella spin structure, depending on the angle $\theta$ formed by the Co spin moments and the axis perpendicular to the plane: $\theta=0$ corresponds to the ferromagnetic order, $\theta=90^{\circ}$ corresponds to the in-plane $120^{\circ}$ spin order. (b) Top and (c) side-top view of the umbrella structure with the notations of the Co sites. }
\label{fig.GGAumbrella}
\end{figure}

 Fig.~\ref{fig.GGAumbrella} shows the results of another constrained GGA calculation, where the magnetic moments at the Co sites were forced to form the ``umbrella structure'', which is characterized by the rotation of spins away from the FM axis $z$ by the angle $\theta$, such that the projections of spins onto the $xy$ plane would form the $120^{\circ}$ structure. Meanwhile, the size of the magnetic moment was allowed to relax during the self-consistency.
\noindent
\noindent The total energy minimum is realized at $\theta =0$, thus confirming that the ground state is ferromagnetic. Then, the total energy gradually increases for $\theta \lesssim 60^{\circ}$ and becomes practically constant afterwards. For $\theta \lesssim 30^{\circ}$, the size of the local magnetic moments at each of the three Co sites, $m_{\nu} = \sqrt{(m_{\nu}^{x})^{2} + (m_{\nu}^{y})^{2} + (m_{\nu}^{z})^{2}}$ ($\nu=$ $1$, $2$, and $3$), is of the order of $0.35$ $\mu_{\rm B}$ and only weakly depends on $\theta$. However, further increase of $\theta$ leads to the collapse of magnetisation: $m_{\nu}$ decreases and becomes equal to zero around $\theta \lesssim 60^{\circ}$ where the total energy reaches the saturation and does not depend on $\theta$, i.e. contrary to what could be expected for localized spins. This is another signature of itineracy of Co$_3$Sn$_2$S$_2$: although the Heisenberg model, which is typically used for the description of localized spins, can still be defined locally, for small rotations of magnetic moments near the FM ground state (as will become evident in Sec.~\ref{sec:Jij}), it breaks down globally, for arbitrary rotations of the magnetic moments by arbitrary angles. Such behavior is not new for the itinerant electron systems: for instance, it is well known that finite rotations of magnetic moments in fcc Ni away from the FM ground state also lead to the collapse of magnetism~\cite{Turzhevskii,Singer}. A similar behaviour is observed in SrRu$_2$O$_6$ and AgRuO$_3$ within GGA, where the sublattice magnetization vanishes upon gradual rotation of spins from the N\'eel AFM ground state to the FM state~\cite{Streltsov,Schnelle2021}.

\par Furthermore, these calculations rule out the existence of $120^{\circ}$ planar structure in Co$_3$Sn$_2$S$_2$, which was proposed to explain the magnetic behavior of Co$_3$Sn$_2$S$_2$ in the temperature range $90$ K $< T <$ $177$ K~\cite{Guguchia.NatComm}, because this $120^{\circ}$ structure does not seem to be compatible with the itinerant character of Co$_3$Sn$_2$S$_2$ as it evolves to the nonmagnetic state.

\par Finally, we would like to emphasize that these calculations were performed without spin-orbit coupling, where the $\theta$-dependence of ${\cal E}$ stems solely from isotropic interactions in the system. It should not be confused with the easy-axis FM anisotropy considered, for instance, in Ref.~\cite{OzawaNomura}.

\section{\label{sec:model} Realistic modelling}
\par In order to estimate the exchange parameters and investigate the stability of the FM state, the electronic states close to the Fermi level were reexpanded in the basis of Wannier functions constructed for the Co $3d$, Sn $5p$, and S $4p$ orbitals using the maximally localized Wannier functions technique~\cite{wannier90}, as described above.

\subsection{\label{sec:existence} Emergence and stability of the magnetic order}
\par First, let us discuss the emergence of the FM state. For these purposes we start with the nonmagnetic solution and evaluate analytically the second derivative of the total energy with respect to the small induced magnetization $\vec{\boldsymbol{m}}$. Very generally, the nonmagnetic state is expected to be unstable because of the kagome flat bands located near the Fermi level~\cite{Yanagi}.

\par In our notations, $\boldsymbol{m}_{\nu} = (m_{\nu}^{x},m_{\nu}^{y},m_{\nu}^{z})$ is the spin magnetic moment at the unit cell site $\nu$, $\vec{\boldsymbol{m}}$ is the column vector assembled from all such $\boldsymbol{m}_{\nu}$ within the unit cell, and $\vec{\boldsymbol{m}}^{T}$ is the row vector corresponding to $\vec{\boldsymbol{m}}$. Then, we formulate the problem in the spirit of constrained spin-density functional theory, where the size and direction of $\vec{\boldsymbol{m}}$ is controlled by the external field $\vec{\boldsymbol{h}}$. The corresponding total energy (per one unit cell) is given by~\cite{PRB2021}
\noindent
\begin{equation}
{\cal E} = {\cal E}_{\rm sp}(\vec{\boldsymbol{h}} + \vec{\boldsymbol{b}}) - \frac{1}{2}\vec{\boldsymbol{m}}^{T} \cdot\left( \vec{\boldsymbol{h}} + \vec{\boldsymbol{b}} \right) + {\cal E}_{\rm xc}(\vec{\boldsymbol{m}}),
\label{eq:tenergy}
\end{equation}
\noindent where the first term is the sum of occupied KS single particle energies (${\cal E}_{\rm sp}$), the second term is minus interaction energy of $\vec{\boldsymbol{m}}$ with $\vec{\boldsymbol{h}}$ and the exchange-correlation (xc) field $\vec{\boldsymbol{b}} = 2 \frac{\delta {\cal E}_{\rm xc}}{\delta \vec{\boldsymbol{m}}}$, and the third terms is the xc energy (${\cal E}_{\rm xc}$), which is taken in the form~\cite{Gunnarsson}
\noindent
\begin{equation}
{\cal E}_{\rm xc} = -\frac{1}{4} \vec{\boldsymbol{m}}^{T} \cdot \hat{\cal I} \vec{\boldsymbol{m}},
\label{eq:exc}
\end{equation}
\noindent so that $\vec{\boldsymbol{b}} = - \hat{\cal I} \vec{\boldsymbol{m}}$ for the site-diagonal matrix $\hat{\cal I} = [\, \dots \, , { \cal I}_{\nu}, \, \dots \, ] $, where $ { \cal I}_{\nu}$ is the Stoner parameter for an ion of the sort $\nu$. Then, it is straightforward to show that
\begin{equation}
{\cal E} = -\frac{1}{4} \vec{\boldsymbol{m}}^{T} \cdot \vec{\boldsymbol{h}}.
\label{eq:emh}
\end{equation}
\noindent Furthermore, $\vec{\boldsymbol{m}}$ can be related to $\vec{\boldsymbol{h}}$ via the response tensor
\noindent
\begin{widetext}
\begin{equation}
{\cal R}_{\boldsymbol{q}}^{\sigma \sigma'} (ab,cd) = \sum_{ml \boldsymbol{k}} \frac{f_{m \boldsymbol{k}}^{\sigma} - f_{l \boldsymbol{k}+\boldsymbol{q}}^{\sigma'}}{\varepsilon_{m \boldsymbol{k}}^{\sigma} - \varepsilon_{l \boldsymbol{k}+\boldsymbol{q}}^{\sigma'}} (C_{m \boldsymbol{k}}^{a\sigma})^{*}C_{l \boldsymbol{k}+\boldsymbol{q}}^{b\sigma'} (C_{l \boldsymbol{k}+\boldsymbol{q}}^{c\sigma'})^{*}C_{m \boldsymbol{k}}^{d\sigma},
\label{eq:gresponse}
\end{equation}
\end{widetext}
\noindent where $\varepsilon_{m \boldsymbol{k}}^{\sigma}$ are the KS eigenvalues and $C_{l \boldsymbol{k}}^{a\sigma}$ are the eigenvectors in the Wannier basis, the pairs of the orbital indices $ab$ and $cd$ belong to the atomic sites $\mu$ and $\nu$, respectively, and $f_{m \boldsymbol{k}}^{\sigma}$ is the Fermi distribution function. In the nonmagnetic states, the elements of the response tensor ${\cal R}_{\mu \nu}^{\sigma \sigma'}$ do not depend on the spin indices. Then, it holds that  $\vec{\boldsymbol{h}} = \left( \hat{\mathbb R}_{0}^{-1} + \hat{\cal I}  \right)  \vec{\boldsymbol{m}}$, where $\hat{\mathbb R}_{\boldsymbol{q}} \equiv [{\mathbb R}_{\boldsymbol{q}, \mu \nu}] $ and
\noindent
\begin{equation}
{\mathbb R}_{\boldsymbol{q}, \mu \nu}^{\sigma \sigma'} = \sum_{a \in \mu} \sum_{c \in \nu} {\cal R}_{\boldsymbol{q}}^{\sigma \sigma'} (aa,cc).
\label{eq:aresponse}
\end{equation}
\noindent
By substituting it in Eq. (\ref{eq:emh}), one obtains
\noindent
\begin{equation}
{\cal E} = \frac{1}{2} \vec{\boldsymbol{m}}^{T} \cdot \hat{\cal D} \vec{\boldsymbol{m}},
\label{eq:emfm}
\end{equation}
\noindent where
\noindent
\begin{equation}
\hat{\cal D} = -\frac{1}{2} \left( \hat{\mathbb R}_{0}^{-1} + \hat{\cal I} \right).
\label{eq:Dmtrx}
\end{equation}

\par The atomic indices $\mu$ and $\nu$ run over the transition-metal (${\rm T}=$ Co) and ligand (${\rm L}=$ Sn or S) sites. Then, the contributions of the ${\rm L}$ variables can be eliminated by assuming that for each instantaneous configuration of the ${\rm T}$ spin moments, the ones at the ligand sites have sufficient time to reach the equilibrium and, therefore, can be found from the adiabaticity condition $\frac{\partial {\cal E}}{\partial \vec{\boldsymbol{m}}_{\rm L}}  = 0$. In this case, ${\cal E}$ can be written as~\cite{PRB2021}
\noindent
\begin{equation}
{\cal E} = \frac{1}{2} \vec{\boldsymbol{m}}^{T}_{\rm T} \cdot \hat{\widetilde{\cal D}}^{\phantom{T}}_{\rm TT} \vec{\boldsymbol{m}}^{\phantom{T}}_{\rm T},
\label{eq:emtfmt}
\end{equation}
\noindent where
\noindent
\begin{equation}
\hat{\widetilde{\cal D}}_{\rm TT} = \hat{\cal D}_{\rm TT} - \hat{\cal D}_{\rm TL}^{\phantom{-1}}\hat{\cal D}_{\rm LL}^{-1}\hat{\cal D}_{\rm LT}^{\phantom{-1}}.
\label{eq:dTT}
\end{equation}

\par Taking into account the symmetry properties for the matrix elements of $\hat{\widetilde{\cal D}}_{\rm TT}$ connecting three Co sites in the primitive cell, $\widetilde{\cal D}_{11} = \widetilde{\cal D}_{22} = \widetilde{\cal D}_{33}$ and $\widetilde{\cal D}_{12} = \widetilde{\cal D}_{23} = \widetilde{\cal D}_{31}$, one obtains:
\noindent
\begin{eqnarray}
{\cal E} & = & \frac{1}{2} \widetilde{\cal D}_{11} \left( \boldsymbol{m}_{1}^{2} + \boldsymbol{m}_{2}^{2} +  \boldsymbol{m}_{3}^{2} \right) + \nonumber \\
 & & \phantom{\frac{1}{2}} \widetilde{\cal D}_{12}  \left( \boldsymbol{m}_{1} \cdot \boldsymbol{m}_{2} + \boldsymbol{m}_{2} \cdot \boldsymbol{m}_{3} + \boldsymbol{m}_{3} \cdot \boldsymbol{m}_{1} \right). \nonumber
\end{eqnarray}
\noindent Considering the directions of the magnetic moments in the umbrella structure (see Fig.~\ref{fig.GGAumbrella} for the geometry and notations of the Co sites),
\noindent
\begin{eqnarray}
  \boldsymbol{m}_{1} &=& (0, \, \sin \theta, \, \cos \theta) \, m, \nonumber \\
  \boldsymbol{m}_{2} &=& (-          \frac{\sqrt{3}}{2} \sin \theta, \, -\frac{1}{2} \sin \theta, \, \cos \theta) \, m, \nonumber \\
  \boldsymbol{m}_{3} &=& (\phantom{-}\frac{\sqrt{3}}{2} \sin \theta, \, -\frac{1}{2} \sin \theta, \, \cos \theta) \, m, \nonumber
\end{eqnarray}
\noindent one can finally obtain the following expression:
\noindent
\begin{equation}
{\cal E} = \frac{3}{2} \left\{ \widetilde{\cal D}_{11} + (3 \cos^2 \theta -1) \widetilde{\cal D}_{12} \right\} m^2.
\label{eq:umbrella}
\end{equation}

\par Therefore, if $\frac{\partial^2 {\cal E}}{\partial m^2} = 3 \{ \widetilde{\cal D}_{11} + (3 \cos^2 \theta -1) \widetilde{\cal D}_{12} \} >0$, the nonmagnetic state is stable. Otherwise, the system will converge to a magnetic solution with finite $m$. In the simplest case of one site in the unit cell, $\hat{\mathbb R}_{0} = -{\cal N}(\varepsilon_{\rm F})$ (the density of states at the Fermi level per one spin) and we recover the conventional criterium of Stoner ferromagnetism: ${\cal I}{\cal N}(\varepsilon_{\rm F})>1$, which can be readily obtained from the condition ${\cal D} <0$ in Eq.~(\ref{eq:emfm}) and using Eq.~(\ref{eq:Dmtrx}) for ${\cal D}$. As expected, the result depends on temperature $T$, which enters this Stoner-type model via the Fermi distribution functions $f_{m \boldsymbol{k}}^{\sigma}$ in Eq.~(\ref{eq:gresponse}). The magnetic structure is stable when $\theta$ is smaller than a certain critical value
\noindent
\begin{equation}
\theta_{m} = \cos^{-1} \sqrt{ \frac{1}{3} \left( 1 - \frac{\widetilde{\cal D}_{11}}{\widetilde{\cal D}_{12}}  \right) }
\label{eq:thetam}
\end{equation}
\noindent for which $\frac{\partial^2 {\cal E}}{\partial m^2} = 0$.

\par
\begin{figure}[t]
\begin{center}
\includegraphics[width=0.48\textwidth]{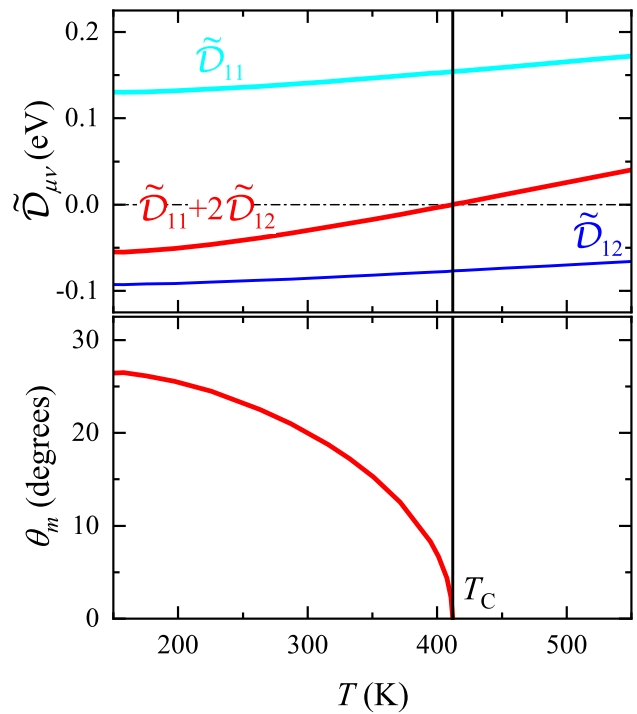}
\end{center}
\caption{
Parameters of the Stoner-type model and temperature dependence of the critical angle of the umbrella structure.}
\label{fig.Stoner}
\end{figure}

 We evaluate these dependencies using the model parameters derived within GGA. In order to obtain the Stoner parameters, ${\cal I}_{\nu} = -\frac{m^{z}_{\nu}}{b^{z}_{\nu}}$, one should know the xc-field $\vec{b}^{z}$ for the given magnetization $\vec{m}^{z}$. It can be obtained from the sum rule $\vec{m}^{z} = \hat{\mathbb R}_{0}^{\uparrow \downarrow} \vec{b}^{z}$~\cite{PRB2021}. Since Eq.~(\ref{eq:exc}) is an approximation, these Stoner parameters depend on the magnetization. Then, for the perturbation theory near the nonmagnetic state, which we consider here, it is logical to derive $\hat{\cal I}$ from the constraint FM calculations with small $M$. More specifically, we use $M = 0.55$ $\mu_{\rm B}$, which yields the following parameters: ${\cal I}_{\rm Co} = 0.97$, ${\cal I}_{\rm Sn_{1}} = -3.52$, ${\cal I}_{\rm Sn_{2}} = -4.65$, and ${\cal I}_{\rm S} = 1.40$ eV. The value of ${\cal I}_{\rm Co}$ is quite consistent with previous estimates for the transition metals~\cite{Gunnarsson}. ${\cal I}_{\rm S}$ is expected to be even larger, as is also known for the oxygen atoms~\cite{MazinSingh1997}. It may look unphysical that ${\cal I}_{\rm Sn_{1}}$ and ${\cal I}_{\rm Sn_{2}}$ are largely negative. However, the small magnetic moments at Sn sites are solely induced by the hybridization with other sites and do not play a primary role in the magnetism of Co$_3$Sn$_2$S$_2$. The response tensor in the nonmagnetic state $\hat{\mathbb R}_{0}$ was evaluated on the mesh of $56 \times 56 \times 56$ $\boldsymbol{k}$-points in the rhombohedral Brillouin zone, which provides a sufficient accuracy at least for $T \gtrsim 150$ K. The results are summarized in Fig.~\ref{fig.Stoner}.
\noindent

\par In the FM state for $\theta = 0$, $\frac{\partial^2 {\cal E}}{\partial m^2} = 3 \{ \widetilde{\cal D}_{11} + 2 \widetilde{\cal D}_{12} \}$ changes the sign around $T_{\rm C}=410$ K, which can be regarded as the Curie temperature of the Stoner model, provided that the transition is not metamagnetic~\cite{Shimizu1}, where $T_{\rm C}$ should be evaluated differently. At the phenomenological level, the conventional practice for the thermodynamic properties of itinerant magnets is to use the Landau-type theory, by expressing the free energy in terms of even powers of $M$: ${\cal E }(M) = \sum_{n=1}^{n_{max}} \frac{1}{2n} a_{2n-1}M^{2n}$ and incorporating the temperature dependence to $a_{1}$ as $a_{1} \to a_{1}\frac{T_{\rm C}-T}{T_{\rm C}}$~\cite{Shimizu2,Mohn}. For instance, for $n_{max}=3$, the metamagnetism occurs if $a_{5}>0$ but $a_{3}<0$. At $T=0$, ${\cal E}(M)$ can be derived from constrained spin density-functional calculations. However, for the half-metallic systems, the dependence ${\cal E}(M)$ is not smooth (see Fig.~\ref{fig.EM}) and such expansion does not apply. Thus, at the moment it is not clear how to proceed in this direction. In any case,  $T_{\rm C}=410$ K can be probably regarded as a rough (an order of magnitude) estimate for the Curie temperature, which exceeds the experimental value by factor 3, as expected for the Stoner-type picture~\cite{Gunnarsson,MoriyaKawabata}.

\par The umbrella structure can be realized for not too large $\theta$ near the FM state. We confirm that there is a critical $\theta_{m}$, which decreases with $T$, and the rotation of magnetic moments beyond this angle makes the umbrella structure unstable relative to the nonmagnetic states, in semi-quantitative agreement with the results of constrained GGA calculations considered in Sec.~\ref{sec:DFT_summary}. Particularly, the critical angle $\theta_{m} \sim 26^{\circ}$ obtained in this model analysis at $T \sim 150$ K is quite consistent with $\theta_{m} \sim 30^{\circ}$ derived from GGA at $T = 0$.

\par Another important point is that $T_{\rm C}$ is expected to decrease with the increase of $\theta$ in the umbrella structure, which immediately follows from Eq.~(\ref{eq:umbrella}) for $\widetilde{\cal D}_{12} < 0$ (see Fig.~\ref{fig.Stoner}). The correspondent dependence $T_{\rm C}(\theta)$ is obtained by inverting the graph $\theta_{m}(T)$, which is also displayed in Fig.~\ref{fig.Stoner}. Thus, the realization of such umbrella structure instead of the collinear FM state could probably rationalize the discrepancy with the experimental data regarding the value of $T_{\rm C}$. Even within the simple Stoner-type picture, considered above, the canting of magnetic moment by $\theta \sim 26^{\circ}$ would be sufficient to produce the experimental $T_{\rm C} \sim 170$ K. Since the nearest Co sites in the Co$_3$Sn$_2$S$_2$ structure are not connected by the inversion symmetry, such canting could be caused by Dzyalishinskii-Moriya interactions~\cite{Dzyaloshinskii_weakF,Moriya_weakF}. This is clearly seen in GGA calculations with the spin-orbit coupling at $T=0$. However, the obtained $\theta$ is too small (only about $2^{\circ}$). It is an interesting question whether $\theta$ will increase with the increase of $T$.

\subsection{\label{sec:Jij} Interatomic exchange interactions in the ferromagnetic state}
\par As we have seen above, finite rotations of spins in Co$_3$Sn$_2$S$_2$ result in the collapse of the magnetic state and in the break down of the Heisenberg model of magnetism. Nevertheless, one can define the model for infinitesimal rotations of spin magnetic moments near the FM ground state. In this section, we construct such model,
\noindent
\begin{equation}
{\cal E} = -\frac{1}{2N} \sum_{ij} J^{ij} \boldsymbol{e}_{i} \cdot \boldsymbol{e}_{j},
\label{eq:Heis}
\end{equation}
\noindent where $\boldsymbol{e}_{i}$ is the direction of spin at the $i$th Co site and $N$ is the number of such sites.

\par For these purposes we consider two techniques. The first one is the standard MFT, which assumes that infinitesimal rotations of spin magnetic moments induce the rotations of the xc fields by the same angles, and this change of the xc fields is treated as a perturbation~\cite{LKAG1987}. The corresponding parameters of exchange interactions between the sublattices $\mu$ and $\nu$ can be found in the reciprocal ($\boldsymbol{q}$) space as
\noindent
\begin{equation}
J^{\mu \nu}_{\boldsymbol{q}} = - \frac{1}{2} \left( b^{z}_{\mu} \left[ \boldsymbol{\mathcal{R}}_{\boldsymbol{q}}^{\uparrow \downarrow} \right]_{\mu \nu} \, b^{z}_{\nu} - b^{z}_{\mu} m^{z}_{\nu} \delta_{\mu \nu} \right).
\label{eq:jmft}
\end{equation}
\noindent In the conventional implementation of MFT, $b^{z}$ and $m^{z}$ are the matrices in the subspace of orbital indices and Eq.~(\ref{eq:jmft}) implies the summation over orbital indices of $b^{z}$, $m^{z}$, and $\boldsymbol{\mathcal{R}}_{\boldsymbol{q}}^{\uparrow \downarrow} \equiv [ {\cal R}_{\boldsymbol{q}}^{\uparrow \downarrow} (ab,cd) ]$. The details can be found in Ref.~\cite{PRB2021}.

\par Nevertheless, MFT is an approximation, which becomes exact only in the long wavelength and strong-coupling limits. However, for the analysis of the exchange interactions, it is essential to go beyond the long wavelength limit and consider the contributions of all $\boldsymbol{q}$ points in the first Brillouin zone. Furthermore, the strong-coupling limit is far from being realized in Co$_3$Sn$_2$S$_2$, as is clearly seen from small values of magnetic moments at the Co sites. Therefore, we consider another technique, which is formally exact as it goes beyond the long wavelength and strong-coupling limits~\cite{BrunoPRL2003}. The corresponding exchange parameters can be found as~\cite{PRB2021}
\noindent
\begin{equation}
J^{\mu \nu}_{\boldsymbol{q}} = \frac{1}{2} \left( m^{z}_{\mu} \left[ {\mathbb R}_{\boldsymbol{q}}^{\uparrow \downarrow} \right]^{-1}_{\mu \nu}  m^{z}_{\nu} - b^{z}_{\mu} m^{z}_{\nu} \delta_{\mu \nu} \right),
\label{eq:jexactM}
\end{equation}
\noindent where $m^{z}_{\mu}$ ($b^{z}_{\mu}$) is the regular (scalar) magnetization (exchange field) at site $\mu$. Similar to MFT, one can also introduce the matrix analog of this expression with $m^{z}_{\mu}$ and $b^{z}_{\mu}$ being the matrices in the subspace of orbital indices. However, the microscopic processes underlying such extension (and describing the rigid rotations of the full magnetization matrix by the same angle) would correspond to much larger energy change, and do not properly capture the low-energy excitations in the system of spins~\cite{PRB2021}.

\par Then, one can start with the bare interactions between the Co sites, which are given by $J^{\mu \nu}_{\boldsymbol{q}}$, and take into account the contributions of the ligand states~\cite{PRB2021}, similar to what we did in Sec.~\ref{sec:existence} in order to understand the emergence of the FM state. The corresponding exchange parameters are given by
\noindent
\begin{equation}
\tilde{J}_{\boldsymbol{q}}^{\rm TT} = J_{\boldsymbol{q}}^{\rm TT} -  J_{\boldsymbol{q}}^{\rm TL} \left[ J_{\boldsymbol{q}}^{\rm LL} \right]^{-1} J_{\boldsymbol{q}}^{\rm LT}.
\label{eq:jTT}
\end{equation}

\par
\begin{figure}[t]
\begin{center}
\includegraphics[width=0.48\textwidth]{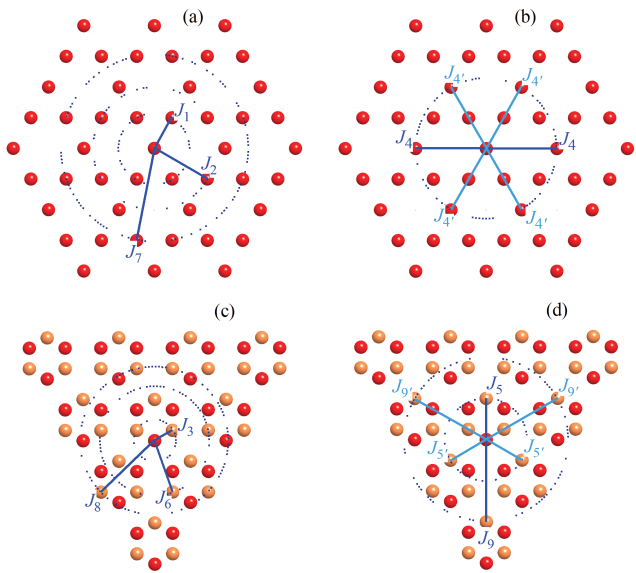}
\end{center}
\caption{
(a), (b) Parameters of interatomic exchange interactions operating in the plane. (c), (d) Parameters operating between the planes (top view).  The Co atoms located in adjacent planes are denoted by different colors. The coordination spheres of atoms around the origin are denoted by dotted circles. (a), (c) Parameters, which have the same value in all bonds for the given coordination sphere. (b), (d) Parameters, which are characterized by two distinct values for two types of inequivalent bonds for the given coordination sphere. The distribution of parameters around two other Co sites in the primitive cell are obtained by the symmetry operation of the space group $R\overline{3}m$.}
\label{fig.Jstr}
\end{figure}

\begin{figure}[t]
\begin{center}
\includegraphics[width=0.48\textwidth]{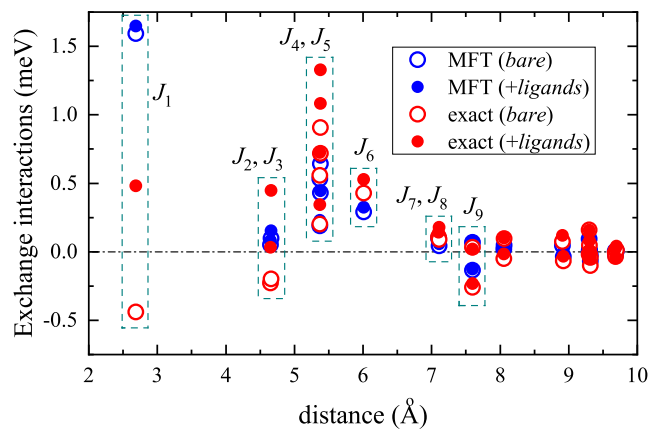}
\end{center}
\caption{
Distance dependence of interatomic exchange interactions as obtained using the magnetic force theorem (MFT) and exact approach: bare parameters of Co-Co interactions and the ones taking into account the contributions of the ligand states. The notations of parameters are explained in Fig.~\ref{fig.Jstr}.}
\label{fig.Jk}
\end{figure}

 Finally, $J^{\mu \nu}_{\boldsymbol{q}}$ and $\tilde{J}^{\mu \nu}_{\boldsymbol{q}}$ can be Fourier transformed to the real space. In these calculations we used the meshes of $40 \times 40 \times 40$ $\boldsymbol{k}$-points and $12 \times 12 \times 12$ $\boldsymbol{k}$-points in the rhombohedral Brillouin zone. Quite expectedly for itinerant systems, the obtained exchange parameters appear to be very long ranged so that sizable interactions can be found even beyond 9th coordinations sphere (Figs.~\ref{fig.Jstr} and \ref{fig.Jk}).
\noindent
\noindent
\noindent Furthermore, the exchange interactions depend on the method, which is used for their calculations, and additional approximations~\cite{PRB2021}. In MFT, the nearest-neighbor interactions in the plane are clearly the strongest (see Fig.~\ref{fig.Jk} and Table~\ref{tab:J}), in agreement with previous studies~\cite{ZhangPRL,Liu2021}. Besides Eq.~(\ref{eq:jmft}), we have also considered a more conventional real-space implementation of MFT based on Green's function technique~\cite{LKAG1987} and confirmed that it provides essentially the same parameters of interatomic exchange interactions.
\noindent
\begin{table}[b]
\caption{Parameters of interatomic exchange interactions (in meV) as obtained in the framework of magnetic force theorem (MFT) and exact formalism: bare Co-Co interactions and the ones taking into account the contributions of the ligand states. The notations of parameters are explained in Fig.~\ref{fig.Jstr}.}
\label{tab:J}
\begin{ruledtabular}
\begin{tabular}{ccccc}
                              & \multicolumn{2}{c}{MFT}    & \multicolumn{2}{c}{exact}      \\
 \cline{2-3}        \cline{4-5}
                              & $bare$    & $+ligands$     & $bare$   & $+ligands$          \\
\hline
$J_{1}$   & $\phantom{-}1.59$    & $\phantom{-}1.65$       & $-0.44$             & $\phantom{-}0.48$       \\
$J_{2}$   & $\phantom{-}0.05$    & $\phantom{-}0.08$       & $-0.23$             & $\phantom{-}0.03$       \\
$J_{3}$   & $\phantom{-}0.10$    & $\phantom{-}0.15$       & $-0.20$             & $\phantom{-}0.45$       \\
$J_{4}$   & $\phantom{-}0.19$    & $\phantom{-}0.23$       & $\phantom{-}0.20$   & $\phantom{-}0.35$       \\
$J_{4}'$  & $\phantom{-}0.53$    & $\phantom{-}0.55$       & $\phantom{-}0.56$   & $\phantom{-}0.72$       \\
$J_{5}$   & $\phantom{-}0.43$    & $\phantom{-}0.45$       & $\phantom{-}0.72$   & $\phantom{-}1.33$       \\
$J_{5}'$  & $\phantom{-}0.64$    & $\phantom{-}0.69$       & $\phantom{-}0.91$   & $\phantom{-}1.08$       \\
$J_{6}$   & $\phantom{-}0.29$    & $\phantom{-}0.33$       & $\phantom{-}0.43$   & $\phantom{-}0.53$       \\
$J_{7}$   & $\phantom{-}0.10$    & $\phantom{-}0.11$       & $\phantom{-}0.11$   & $\phantom{-}0.14$       \\
$J_{8}$   & $\phantom{-}0.04$    & $\phantom{-}0.06$       & $\phantom{-}0.09$   & $\phantom{-}0.18$       \\
$J_{9}$   & $-0.13$              & $-0.13$                 & $-0.26$             & $-0.23$                 \\
$J_{9}'$  & $\phantom{-}0.07$    & $\phantom{-}0.08$       & $\phantom{-}0.03$   & $\phantom{-}0.02$
\end{tabular}
\end{ruledtabular}
\end{table}
\noindent The contributions of the ligand states in this case are relatively unimportant and the main tendencies of $J_{ij}$ are captured already by the bare interactions between Co sites.

\par This picture changes significantly in the exact approach, where the strongest interaction is $J_{5}$ in the 5th coordination sphere (see Fig.~\ref{fig.Jstr}). Since $J_{5}$ operates between the planes, Co$_3$Sn$_2$S$_2$ in our picture is essentially a three-dimensional material. Furthermore, the ligands states appear to be very important in this case, as they strengthen the FM character of interactions and are primarily responsible for the FM origin of these interactions in the first three coordination spheres. Nevertheless, $T_{\rm C}$ evaluated in the Heisenberg model appears to be smaller than the experimental one. Particularly, the molecular field approximation, where $T_{\rm C} = \frac{1}{3k_{\rm B}} \sum_{j}J_{0j}$, is known to overestimate $T_{\rm C}$. However, if we applied this approximation to Co$_3$Sn$_2$S$_2$, we would get only $T_{\rm C}=$ $77$ and $95$ K in framework of MFT and the exact approach, respectively.

\par The results of recent inelastic neutron scattering data were interpreted in terms of three parameters~\cite{ZhangPRL}: $J_{2} = -0.08$, $J_{c1} = 0.44$, and $J_{d} = 0.81$ meV (corresponding to $J_{2}$, $J_{3}$, and $J_{4}$ in our notations)~\cite{footnote1}. Thus, the strongest interaction is expected to be $J_{4}$ (the so-called ``cross-hexagon'' interaction), while the nearest-neighbor coupling $J_{1}$ is neglegibly small. This interpretation is clearly inconsistent with theoretical calculations based on MFT, where $J_{1}$ is the strongest. Nevertheless, there is also a considerable difference from the results of the exact method, where the strongest interaction is $J_{5}$ ($J_{c3}$ in the notations of Ref.~\cite{ZhangPRL}), while the ``cross-hexagon'' interaction $J_{4}$ is substantially smaller. Furthermore, even within the 4th coordination sphere, $J_{4}$ is not the strongest interaction and $J_{4}'$ is considerably stronger than $J_{4}$. It is also interesting to note that $J_{4}$ and $J_{5}$ operate practically at the same distances: $J_{4}$ is within the plane, while $J_{5}$ is between the planes (see Fig.~\ref{fig.Jk}).

\par
\begin{figure}[b]
\begin{center}
\includegraphics[width=0.48\textwidth]{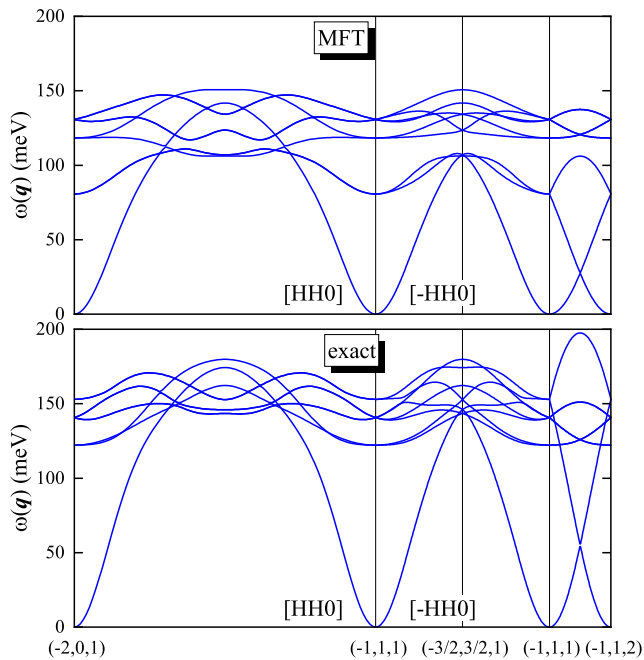}
\end{center}
\caption{
Spin-wave dispersion corresponding to the exchange parameters derived in the framework of magnetic force theorem (MFT) and exact approach. All notations are taken from Ref.~\cite{ZhangPRL} for the hexagonal lattice.}
\label{fig.SWhex}
\end{figure}

 We hope that the results of our theoretical calculations of interatomic exchange interactions could be used as the guideline for the interpretation of experimental inelastic neutron scattering data. In Fig.~\ref{fig.SWhex}, we plot the theoretical spin-wave dispersion, which was defined as eigenvalues $\omega_{n \boldsymbol{q}}$ of the $3 \times 3$ matrix $\hat{\Omega}_{\boldsymbol{q}} = [\Omega_{\boldsymbol{q}}^{\mu \nu}]$ (for 3 magnetic Co sublattices in the rhombohedral unit cell or the $9 \times 9$ matrix for the hexagonal cell including 9 Co atoms), where
\noindent
\begin{equation}
\Omega_{\boldsymbol{q}}^{\mu \nu} = \frac{2}{m}\left( J^{\mu} \delta_{\mu \nu} - J^{\mu \nu}_{\boldsymbol{q}} \right),
\label{eqn.SW}
\end{equation}
\noindent $J^{\mu \nu}_{\boldsymbol{q}}$ is the Fourier image of $J_{ij}$ between sublattices $\mu$ and $\nu$, and $J^{\mu} = \sum_{\nu} J^{\mu \nu}_{0}$.
\noindent
\noindent We consider the results based on the MFT and exact technique, taking into account the contributions of the ligand states, and use the notations of Ref.~\cite{ZhangPRL} for the hexagonal lattice. In fact, the experimental spin-wave dispersion was measured only for not too large values of $\boldsymbol{q}$ around the $\Gamma$ point, which is denoted as $(-1,1,1)$ in Fig.~\ref{fig.SWhex}, and limited only to the acoustic ($A$) branch. The key feature of this experimental data is that the magnon dispersion along the $[HH0]$ direction is considerably steeper than the one along $[-HH0]$. This anisotropy of the magnon spectrum was suggested to be the main signature of the strong ``cross-hexagon'' interaction $J_{4}$, as other theoretical models used for the fitting of the experimental data led to very similar dispersion along the $[HH0]$ and $[-HH0]$ directions~\cite{ZhangPRL}. Nevertheless, this explanation looks disputable in the light of the following arguments: the behavior of the $A$ branch near the $\Gamma$ point is described by the spin-stiffness tensor $\hat{D} = [D^{\alpha \beta}]$:
\noindent
\begin{equation}
\omega_{L \boldsymbol{q}} = \sum_{\alpha, \beta} D^{\alpha \beta} q_{\alpha} q_{\beta},
\label{eqn.DSW}
\end{equation}
\noindent where $\alpha, \beta=$ $x$, $y$, or $z$. For the $R$-$3m$ symmetry, $\hat{D}$ is diagonal and $D^{xx}=D^{yy}$. Thus, the spin-wave dispersion near the $\Gamma$ point caused by isotropic exchange interactions, including the ``cross-hexagon'' $J_{4}$, must be isotropic in the $xy$ plane, and this is exactly what is seen in our calculations in Fig.~\ref{fig.SWhex}.

\par Although the exact approach for the interatomic exchange interactions better captures the total energy change, caused by the infinitesimal rotations of spins, it is believed that MFT is more suitable for the analysis of the spin-wave dispersion~\cite{KL2004}. Nevertheless, in the long wavelength limit $\boldsymbol{q} \to 0$ these two techniques provide very similar description~\cite{BrunoPRL2003,PRB2021}, as is clearly seen in Fig.~\ref{fig.SWhex}, while the main difference occurs in the high-energy region of optical branches.

\par The experimental anisotropy of the spin-wave dispersion in the $xy$ plane is an interesting point~\cite{ZhangPRL}. However, it is probably caused by other mechanisms and not related to isotropic exchange interactions.

\subsection{\label{sec:Jm} Magnetic moments dependence of the exchange interactions}
\par The picture of collinear FM spins, whose size changes with temperature, is at the heart of the Stoner model of magnetism~\cite{Stoner}. Nevertheless, it is reasonable to expect that besides these changes (longitudinal fluctuations), the spins can experience the infinitesimal rotations near the equilibrium state (or transversal fluctuations), which can be regarded as the step towards a more general spin fluctuation theory~\cite{Takahashi,MoriyaKawabata,MoriyaSF,UhlKubler}. In this section, we explore the effect of the size of the magnetic moments on the stability of the FM state with respect to the spin rotational degrees of freedom employing a somewhat phenomenological strategy for these purposes. Namely, we perform constrained GGA calculations, where we additionally fix the value of the total magnetic moment and, then, using the so-obtained constrained electronic structure we evaluate parameters of interatomic exchange interactions. A similar strategy was used for the analysis of photoemission~\cite{HolderPRPphoto} and optical~\cite{YangPRLoptics} data. As for the exchange interactions, we consider here only the exact approach, Eq.~(\ref{eq:jexactM}), and take into account the contributions of the ligand states using Eq.~(\ref{eq:jTT}). In our constraint calculations, we fix the total moment of three Co sites in the unit cell (evaluated within atomic spheres of radii $1.3$~\AA) to $0.32$, $0.53$, and $0.83$ $\mu_{B}$. This corresponds to the following values of total magnetic moments in the unit cell (and including the contributions of the Sn and S sites): $M=$ $0.55$, $0.78$, and $0.99$ $\mu_{B}$, which are considered together with the results of unconstrained calculations with $M=1$ $\mu_{B}$ ($1.035$ $\mu_{B}$ within atomic Co spheres). Particularly, we will show that with the decrease of $M$, the FM state becomes unstable and this instability may be related to the emergence of some AFM phase at elevated $T$, which was observed experimentally in Ref.~\cite{Guguchia.NatComm}. Using the experimental dependence $M(T)$ reported in Ref.~\cite{Kassem}, the values of $M=$ $0.55$, $0.78$, and $0.99$ $\mu_{B}$ can be roughly related to the temperatures $T/T_{\rm C} \sim$ $0.95$, $0.75$, and $0.2$, respectively.

\par
\begin{figure}[t]
\begin{center}
\includegraphics[width=0.48\textwidth]{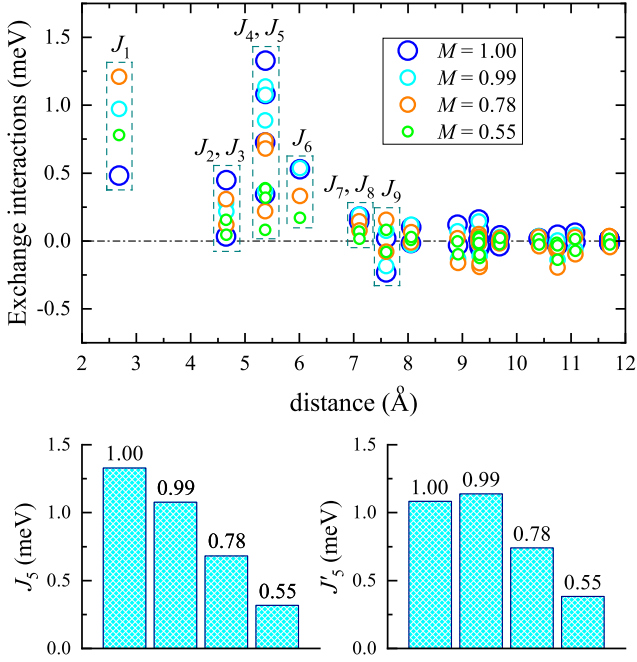}
\end{center}
\caption{
(Top) Distance dependence of interatomic exchange interactions in the exact approach including the contributions of the ligand states as obtained for the constrained electronic structure with fixed values of total magnetic moments $M$. (Bottom) Dependence of interplane interactions $J_{5}$ and $J'_{5}$ on $M$. The notations of parameters are explained in Fig.~\ref{fig.Jstr}.}
\label{fig.Jm}
\end{figure}

 Distance dependence of the exchange interactions for different values of $M$ is shown in Fig.~\ref{fig.Jm}.
\noindent
\noindent Particularly, we note that the decrease of $M$ strengthens the nearest-neighbor interaction $J_{1}$, which gradually starts to dominate over other exchange interactions. On the other hand, the interplane interactions $J_{5}$ and $J'_{5}$ decrease with the decrease of $M$. Moreover, some long-range interplane interactions beyond the 9th coordination sphere become more antiferromagnetic. Thus, one can expect the weakening of the FM coupling between the planes with the decrease of $M$.

\par
\begin{figure}[t]
\begin{center}
\includegraphics[width=0.48\textwidth]{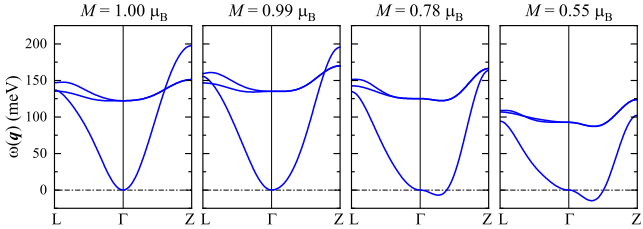}
\end{center}
\caption{
Spin-wave dispersion in the ferromagnetic state for different values of total magnetic moment. The coordinates of the high symmetry points of the rhombohedral Brillouin zone are ${\rm L}: (\frac{\pi}{\sqrt{3}a},\frac{\pi}{3a},\frac{\pi}{3c})$, $\Gamma:(0,0,0)$, and ${\rm Z}:(0,0,\frac{\pi}{c})$.}
\label{fig.SW}
\end{figure}

 Using the obtained exchange parameters, we evaluate the stability of the FM state. For this purpose we calculate the magnon energies, which are given by the eigenvalues of Eq.~(\ref{eqn.SW}) for the rhombohedral lattice. If some of the $\omega_{n \boldsymbol{q}}$'s are negative, the state is unstable for those $\boldsymbol{q}$'s. The results are shown in Fig.~\ref{fig.SW}.
\noindent
\noindent Furthermore, we evaluate the spin-stiffness tensor $\hat{D} = [D^{\alpha \beta}]$ for the $A$ branch. The non-vanishing matrix elements $D^{xx}=D^{yy}$ and $D^{zz}$ of $\hat{D}$ are listed in Table~\ref{tab:D}.
\noindent
\begin{table}[b]
\caption{Matrix elements of the spin-stiffness tensor (in meV/\AA$^2$) for different values of total magnetic moment $M$ (in $\mu_{\rm B}$). The values obtained after excluding the Fermi surface contributions are given in parentheses.}
\label{tab:D}
\begin{ruledtabular}
\begin{tabular}{rrrrr}
 $M$      & \multicolumn{2}{c}{$D^{xx}$}   & \multicolumn{2}{c}{$D^{zz}$}     \\
\hline
$1.00$    & $1019$    & ($1019$)           & $1107$   &  ($1107$)             \\
$0.99$    & $957$     & ($961$)            & $527$    &  ($545$)              \\
$0.78$    & $639$     & ($644$)            & $-421$   &  ($-387$)             \\
$0.55$    & $469$     & ($467$)            & $-565$   &  ($-498$)
\end{tabular}
\end{ruledtabular}
\end{table}
\noindent In the ground state ($M=1$ $\mu_{\rm B}$) the tensor $\hat{D}$ is nearly isotropic ($D^{xx} \approx D^{zz}$). However, even small deviation from the ground state for $M=0.99$ $\mu_{\rm B}$ leads to the sharp drop of $D^{zz}$ and moderate decrease of $D^{xx}$. Such drop is caused by the discontinuity of the electronic structure related to the deviation from the half-metallic state, which also leads to the derivative discontinuity of ${\cal E}(M)$ as shown in Fig.~\ref{fig.EM}. The obtained values are still larger than the experimental $D^{xx} = 803 \pm 46$ and $D^{zz} = 237 \pm 13$ meV/\AA$^2$ measured at $T=8$ K~\cite{Liu2021}. Nevertheless, these parameters are very sensitive to the value of $M$ (and the ordered moment at the Co site, reported in Ref.~\cite{Liu2021}, was smaller than $0.3$ $\mu_{\rm B}$, meaning that the measured sample was probably not in the half-metallic state). Indeed, further decrease of $M$ makes $D^{zz}<0$ and the FM state becomes unstable. $D^{xx}$ also decreases with the decrease of $M$, but remains positive for all considered values of $M$. Such instability is resolved in the formation of an incommensurate spin-spiral state with $\boldsymbol{q}=(0,0,q_{\rm z})$ as confirmed by the spin-wave calculations in Fig.~\ref{fig.SW}.

\par In order to study the effect of the Fermi surface contributions to the exchange parameters, we eliminate these contributions by enforcing $\frac{f_{m \boldsymbol{k}}^{\sigma} - f_{l \boldsymbol{k}+\boldsymbol{q}}^{\sigma'}}{\varepsilon_{m \boldsymbol{k}}^{\sigma} - \varepsilon_{l \boldsymbol{k}+\boldsymbol{q}}^{\sigma'}} = 0$ in Eq.~(\ref{eq:gresponse}) for $\varepsilon_{m \boldsymbol{k}}^{\sigma} \rightarrow \varepsilon_{l \boldsymbol{k}+\boldsymbol{q}}^{\sigma'}$. Although the effect of the Fermi surface states on the individual exchange interactions does not look strong, there is an appreciable contribution of these states to the spin stiffness, mainly associated with the long-range interactions. The results are given in parentheses in Table~\ref{tab:D}. As expected, there is no Fermi surface contribution to $\boldsymbol{\mathcal{R}}_{\boldsymbol{q}}^{\uparrow \downarrow}$ in the half-metallic ground state with $M = 1$ $\mu_{\rm B}$. In the metallic state with $M < 1$ $\mu_{\rm B}$, the contribution of the Fermi surface states to $D^{xx}$ is negligibly small. Nevertheless, there is an appreciable AFM contribution of the surface states to $D^{zz}$, which additionally destabilizes the FM state.

\par Thus, we expect that with the increase of $T$, when the magnetic moments become sufficiently small, Co$_3$Sn$_2$S$_2$ can undergo the transition to the incommensurate AFM state. At present, we cannot elaborate details of this transition (for instance, whether it goes via the region of coexistence of the FM and AFM phases). Nevertheless, we believe that such behavior may be relevant to the anomalous properties of Co$_3$Sn$_2$S$_2$ for $T > 90$ K~\cite{Guguchia.NatComm}.

\section{\label{sec:summary} Summary and Conclusions}
\par Using results of density functional theory in the generalized gradient approximation, we investigated the origin and stability of the FM order in the Weyl semimetal Co$_3$Sn$_2$S$_2$. For these purposes, we constructed the realistic model in the basis of localized Wannier functions, which included the contributions of the Co $3d$ as well as the ligand Sn $5p$, and S $4p$ states, and studied this model using different types of the response theories.

\par One of the interesting aspects of Co$_3$Sn$_2$S$_2$ is that the local magnetic moments are rather soft and strongly depend on the angle formed by the Co spins in the kagome lattice. This is one of the key results of magnetic GGA calculations, which is nicely reproduced by the response theory, by considering the emergence of the magnetic solutions starting from the nonmagnetic state. This finding strongly supports the itinerant character of magnetism in Co$_3$Sn$_2$S$_2$, which should be considered in the analysis of properties of this compound. For instance, the size of the local magnetic moments should depend on temperature, which should be one of the genuine physical properties of Co$_3$Sn$_2$S$_2$.

\par On the other hand, the Heisenberg model of localized magnetism also makes sense in the case of Co$_3$Sn$_2$S$_2$ for the analysis of local stability of the FM state with respect to the transversal spin fluctuations, inherent to rotational spin degrees of freedom. For the construction of such model, we employed the exact theory of interatomic exchange interactions based on the calculation of the inverse response function. We argued that the interatomic exchange interactions in Co$_3$Sn$_2$S$_2$ are very long-ranged and the strongest one, stabilizing the FM state, is operating in the 5th coordinate spheres between the kagome planes.

\par Furthermore, we expect the FM magnetization to decrease with temperature via the longitudinal fluctuations, affecting the size of magnetic moments. This will destroy the half-metallic character of Co$_3$Sn$_2$S$_2$ and gradually makes the FM state unstable with respect to the transversal fluctuations. The change of the electronic structure mainly affects the interactions between the kagome planes, partly owing to the contributions stemming from the Fermi surface states. Thus, with the increase of $T$, we expect Co$_3$Sn$_2$S$_2$ to change gradually from a three-dimensional to quasi-two-dimensional ferromagnet, which should be followed by emergence of the spin-spiral phase propagating perpendicular to the kagome planes. This finding could probably rationalize the experimental behavior of Co$_3$Sn$_2$S$_2$ near $T_{\rm C}$~\cite{Guguchia.NatComm}.

\par Another important question is the validity of GGA, which is typically employed for the analysis of Weyl semimetal properties of Co$_3$Sn$_2$S$_2$. From the viewpoint of interatomic exchange interactions, the experimental information available on hands is not sufficient to make a definite conclusion. The anisotropy of the spin stiffness, which is measured experimentally~\cite{ZhangPRL,Liu2021}, can be understood by a small deviation from the half-metallic state. We hope that the comprehensive analysis presented in our work can be used as the guideline for future experimental studies. Particularly, it would be interesting to check our finding that the strongest exchange interaction stabilizing the FM state in Co$_3$Sn$_2$S$_2$ operates in the 5th coordination sphere, between the kagome plane. The theory for $T_{\rm C}$ in Co$_3$Sn$_2$S$_2$ should involve the aspects of both Stoner and Heisenberg theories of magnetism~\cite{Takahashi,MoriyaKawabata,MoriyaSF,UhlKubler}. Separately, none of these models would provide a reasonable description for Co$_3$Sn$_2$S$_2$.

\par In the present work, we had to deal with the extended model in the basis of Co $3d$, Sn $5p$, and S $4p$ states, similar to the previous studies~\cite{Yanagi,Minami}. A very interesting direction is the formulation of effective toy theories for magnetic Weyl semimetals, which would capture the behavior of small number states near the Fermi level~\cite{OzawaNomura}. Although this can be done rigorously by employing the Wannier function technique~\cite{wannier90}, such a construction for Co$_3$Sn$_2$S$_2$ and similar materials is not always straightforward because of the clustering effects and the formation of molecular groups of states~\cite{SavrasovPRB2021}.

\section*{Acknowledgement}
\par IVS acknowledges useful communication with V. P. Antropov, drawing our attention to Refs.~\cite{Turzhevskii,Singer} on validity of the Heisenberg model for transition metals, and S. Okamoto on details of Ref.~\cite{ZhangPRL}. The work was supported by program AAAA-A18-118020190095-4 (Quantum).


\begin{thebibliography}{99}

\bibitem{Liu2}
D.~F. Liu, A.~J. Liang, E.~K. Liu, Q.~N. Xu, Y.~W. Li, C. Chen, D. Pei, W.~J. Shi, S.~K. Mo, P. Dudin, T. Kim, C. Cacho, G. Li, Y. Sun, L.~X. Yang, Z.~K. Liu, S.~S.~P. Parkin, C. Felser, and Y.~L. Chen, Science \textbf{365}, 1282 (2019).

\bibitem{Xu}
Q. Xu, E. Liu, W. Shi, L. Muechler, J. Gayles, C. Felser, and Y. Sun,
Phys. Rev. B \textbf{97}, 235416 (2018).

\bibitem{Liu}
E. Liu, Y. Sun, N. Kumar, L. Muechler, A. Sun, L. Jiao, Sh.-Y. Yang, D. Liu, A. Liang, Q. Xu, J. Kroder, V. S\"u\ss, H. Borrmann, Ch. Shekhar, Zh. Wang, Ch. Xi, W. Wang, W. Schnelle, S. Wirth, Y. Chen, S.~T.~B. Goennenwein, and C. Felser,
Nature Phys.  \textbf{14}, 1125 (2018).

\bibitem{Yanagi}
Y. Yanagi, J. Ikeda, K. Fujiwara, K. Nomura, A. Tsukazaki, and M.-T. Suzuki, Phys. Rev. B \textbf{103}, 205112 (2021).

\bibitem{Minami}
S. Minami, F. Ishii, M. Hirayama, T. Nomoto, T. Koretsune, and R. Arita, Phys. Rev. B \textbf{102}, 205128 (2020).

\bibitem{Tanaka}
M. Tanaka, Y. Fujishiro, M. Mogi, Y. Kaneko, T. Yokosawa, N. Kanazawa, S. Minami, T. Koretsune, R. Arita, S. Tarucha, M. Yamamoto, and Y. Tokura, Nano Lett. \textbf{20}, 7476 (2020).

\bibitem{Muechler}
L. Muechler, E. Liu, J. Gayles, Q. Xu, C. Felser, and Y. Sun, Phys. Rev. B \textbf{101}, 115106 (2020).

\bibitem{Jiao}
L. Jiao, Q. Xu, Y. Cheon, Y. Sun, C. Felser, E. Liu, and S. Wirth,
Phys. Rev. B \textbf{99}, 245158 (2019).

\bibitem{Zhou}
H. Zhou, G. Chang, G. Wang, X. Gui, X. Xu, J.-X. Yin, Z. Guguchia, S. S. Zhang, T.-R. Chang, H. Lin, W. Xie, M. Z. Hasan, Sh. Jia, Phys. Rev. B \textbf{101}, 125121 (2020).

\bibitem{Kassem} M. A. Kassem, \textit{PhD Dissertation} (Kyoto Univ., 2016).

\bibitem{Guguchia.NatComm}
Z. Guguchia, J.~A.~T. Verezhak, D.~J. Gawryluk, S.~S. Tsirkin, J.-X. Yin, I. Belopolski, H. Zhou, G. Simutis, S.-S. Zhang, T.~A. Cochran, G. Chang, E. Pomjakushina, L. Keller, Z. Skrzeczkowska, Q. Wang, H.~C. Lei, R. Khasanov, A. Amato, S. Jia, T. Neupert, H. Luetkens, and M.~Z. Hasan, Nat. Commun. \textbf{11}, 559 (2020).

\bibitem{ZhangPRL}
Q. Zhang, S. Okamoto, G.~D. Samolyuk, M.~B. Stone, A.~I. Kolesnikov, R. Xue, J. Yan, M.~A. McGuire, D. Mandrus, and D.~A. Tennant, Phys. Rev. Lett. \textbf{127}, 117201 (2021).

\bibitem{SohArxiv}
J.-R. Soh, Ch.-J. Yi, I. Zivkovic, N. Qureshi, A. Stunault, B. Ouladdiaf, J.~A. Rodr\'iguez-Velamaz\'an, Y.-G. Shi, and A.~T. Boothroyd, arXiv:2110.00475 [cond-mat.str-el].

\bibitem{SavrasovPRB2021}
A. Rossi, V. Ivanov, S. Sreedhar, A.~L. Gross, Z. Shen, E. Rotenberg, A. Bostwick, Ch. Jozwiak, V. Taufour, S.~Y. Savrasov, and I.~M. Vishik, Phys. Rev. B \textbf{104}, 155115 (2021).

\bibitem{lacunar}
S.~A. Nikolaev and I. V. Solovyev, Phys. Rev. B \textbf{99}, 100401(R) (2019);
S.~A. Nikolaev and I. V. Solovyev, Phys. Rev. B \textbf{102}, 014414 (2020).

\bibitem{RMP} M.~I. Katsnelson, V.~Yu. Irkhin, L. Chioncel, A.~I. Lichtenstein, R.~A. de Groot, Rev. Mod. Phys. \textbf{80}, 315 (2008).

\bibitem{Takahashi}
Y. Takahashi, \textit{Spin Fluctuation Theory of Itinerant Electron Magnetism}, in Springer Tracts in Modern Physics, edited by G. H\"ohler, A. Fujimori, J.~H. K\"uhn, T. M\"ller, F. Steiner, W.~C. Stwalley, J.~E. Tr\"umper, P. W\"olfle, U. Woggon (Springer-Verlag, Berlin 2013), Vol. 253.

\bibitem{Sakon}
T. Sakon, Y. Hayashi, A. Fukuya, D. Li, F. Honda, R.~Y. Umetsu, X. Xu, G. Oomi, T. Kanomata, and T. Eto, Materials \textbf{12}, 575 (2019).

\bibitem{kat07}
A.~A. Katanin and V.~Yu. Irkhin, Phys. Usp. \textbf{50}, 613 (2007).

\bibitem{Liu2021}
C. Liu, J.-L. Shen, J.-C. Gao, C.-J. Yi, D. Liu, T. Xie, L. Yang, S. Danilkin, G.-C. Deng, W.-H. Wang, S.-L. Li, Y.-G. Shi, H.-M. Weng, E.-K. Liu, and H.-Q. Luo, Sci. China Phys. Mech. Astron. \textbf{64}, 217062 (2021).

\bibitem{LKAG1987}
A.~I. Liechtenstein, M.~I. Katsnelson, V.~P. Antropov, and V.~A. Gubanov, J. Magn. Magn. Mater. \textbf{67}, 65 (1987).

\bibitem{BrunoPRL2003}
P. Bruno, Phys. Rev. Lett. \textbf{90}, 087205 (2003).

\bibitem{Antropov2006}
V.~P. Antropov, M. van Schilfgaarde, S. Brink, and J.~L. Xu, J. Appl. Phys. \textbf{99}, 08F507 (2006).

\bibitem{PRB2021}
I.~V. Solovyev, Phys. Rev. B \textbf{103}, 104428 (2021).

\bibitem{Stoner}
E.~C. Stoner, Proc. Royal. Soc. A \textbf{154}, 656 (1936).

\bibitem{gga-pbe}
J.~P. Perdew, K. Burke, and M. Ernzerhof, Phys. Rev. Lett. \textbf{77}, 3865 (1996).

\bibitem{structure}
P. Vaqueiro and G.~G. Sobany, Solid State Sciences \textbf{11}, 513 (2009).

\bibitem{vasp}
G.~Kresse and J.~Hafner, Phys. Rev. B \textbf{47}, 558 (1993).

\bibitem{paw}
P.~E. Blochl, Phys. Rev. B  \textbf{50}, 17953 (1994).

\bibitem{mpack}
H.~J.~Monkhorst and J.~D. Pack, Phys. Rev. B \textbf{13}, 5188 (1976).

\bibitem{MethfesselPaxton}
M. Methfessel and A.~T. Paxton, Phys. Rev. B \textbf{40}, 3616 (1989).

\bibitem{wannier90}
A.~A. Mostofi, J.~R. Yates, G. Pizzi, Y.~S. Lee, I. Souza, D. Vanderbilt, and N. Marzari,
Comput. Phys. Commun. \textbf{185}, 2309 (2014).

\bibitem{WannierTools}
Q. Wu, Sh. Zhang, H.-F. Song, M. Troyer, A.~A. Soluyanov. Computer Physics Communications \textbf{224}, 405 (2018).

\bibitem{EschrigPickett}
H. Eschrig and W.~E. Pickett, Solid State Commun. \textbf{118}, 123 (2001).

\bibitem{Turzhevskii}
S.~A. Turzhevskii, A.~I. Lichtenstein, and M.~I. Katsnelson, Fiz. Tverd. Tela \textbf{32}, 1952 (1990) [Sov. Phys. Solid State \textbf{32}, 1138 (1990)].

\bibitem{Singer}
R. Singer, M. F\"ahnle, and G. Bihlmayer, Phys. Rev. B \textbf{71}, 214435 (2005).

\bibitem{Streltsov}
S. Streltsov, I.~I. Mazin, and K. Foyevtsova, Phys. Rev. B \textbf{92}, 134408 (2015).

\bibitem{Schnelle2021}
W. Schnelle, B.~E. Prasad, C. Felser, M. Jansen, E.~V. Komleva, S.~V. Streltsov, I.~I. Mazin, D.~D. Khalyavin, P. Manuel, S. Pal, D.~V.~S. Muthu, A.~K. Sood, E.~S. Klyushina, B. Lake, J.-C. Orain, and H. Luetkens, Phys. Rev. B {\bf 103}, 214413 (2021).

\bibitem{OzawaNomura}
A. Ozawa and K. Nomura, J. Phys. Soc. Jpn. \textbf{88}, 123703 (2019).

\bibitem{Gunnarsson}
O. Gunnarsson, J. Phys. F: Met. Phys. \textbf{6}, 587 (1976).

\bibitem{MazinSingh1997}
I.~I. Mazin and D.~J. Singh, Phys. Rev. B \textbf{56}, 2556 (1997).

\bibitem{Shimizu1}
M. Shimizu, J. Physique \textbf{43}, 155 (1982).

\bibitem{Shimizu2}
M. Shimizu, J. Physique \textbf{43}, 681 (1982).

\bibitem{Mohn}
P. Mohn, \textit{Magnetism in the Solid State}, in Springer Series in solid-state sciences, edited by
M. Cardona, P. Fulde, K. von Klitzing, R. Merlin, H.-J. Queisser, H. St\"ormer (Springer-Verlag, Berlin 2006), Vol. 134.

\bibitem{MoriyaKawabata}
T. Moriya and A. Kawabata, J. Phys. Soc. Jpn. \textbf{34}, 639 (1973);
T. Moriya and A. Kawabata, \textit{ibid.} \textbf{35}, 669 (1973).

\bibitem{Dzyaloshinskii_weakF}
I. Dzyaloshinsky,
J. Chem. Phys. Solids \textbf{4}, 241 (1958).

\bibitem{Moriya_weakF}
T. Moriya,
Phys. Rev. \textbf{120}, 91 (1960).

\bibitem{footnote1} In order to be consistent with our definition, the exchange parameters of Ref.~\cite{Guguchia.NatComm} were additionally multiplied by $2S=0.37$ and transformed from K units to meV.

\bibitem{KL2004}
M.~I. Katsnelson and A.~I. Lichtenstein, J. Phys.: Condens. Matter \textbf{16}, 7439 (2004).

\bibitem{MoriyaSF}
T. Moriya, \textit{Spin Fluctuations in Itinerant Electron Magnetism}
(Springer, Berlin, 1985).

\bibitem{UhlKubler}
M. Uhl and J. K\"ubler, Phys. Rev. Lett. \textbf{77}, 334 (1996).

\bibitem{HolderPRPphoto}
M. Holder, Yu.~S. Dedkov, A. Kade, H. Rosner, W. Schnelle, A. Leithe-Jasper, R. Weihrich, and S.~L. Molodtsov, Phys. Rev. B \textbf{79}, 205116 (2009).

\bibitem{YangPRLoptics}
R. Yang, T. Zhang, L. Zhou, Y. Dai, Zh. Liao, H. Weng, and X. Qiu, Phys. Rev. Lett. \textbf{124}, 077403 (2020).


\end{thebibliography}
\end{document}